\newcounter{myctr}
\def\myitem{\refstepcounter{myctr}\bibfont\noindent\ifnum\themyctr>9\else\phantom{0}\fi\hangindent17pt\themyctr.\enskip}

\documentclass{ws-ijqi}
\newcommand{\be}{\begin{equation}}
\newcommand{\ee}{\end{equation}}
\newcommand{\bea}{\vspace{0.25cm}\begin{eqnarray}}
\newcommand{\eea}{\end{eqnarray}}

\def\PRL{{Phys. Rev. Lett.} }
\def\PRA{{Phys. Rev.} A }
\begin{document}

\markboth{} {}

\catchline{}{}{}{}{}

\title{TOWARD THIRD ORDER GHOST IMAGING WITH THERMAL LIGHT\\}

\author{G. BRIDA, I. P. DEGIOVANNI, G. A. FORNARO, M. GENOVESE, A. MEDA
\\
}

\address{Istituto Nazionale di Ricerca Metrologica (INRIM), Strada delle Cacce 91,\\ Torino 10135, Italy\\
a.meda@inrim.it}

\maketitle

\begin{history}
\received{Day Month Year}
\revised{Day Month Year}
\end{history}

\begin{abstract}
Recently it has been suggested that an enhancement in the visibility
of ghost images obtained with thermal light can be achieved
exploiting higher order correlations \cite{3}. This paper reports on
the status of an higher order ghost imaging experiment carried on at
INRIM labs exploiting a pseudo-thermal source and a CCD camera.

\end{abstract}


\section{Introduction}  

Quantum Imaging is a new quantum technology addressed to exploit
properties of quantum optical states for overcoming limits of
classical optics \cite{l}.

Various protocols have been recently proposed
\cite{th1,th2,th3,th4,th5,th6} and realized
\cite{exp1,exp2,exp3,exp4,exp5,exp6} ranging from super-resolution
to sub-shot noise imaging.

One of the oldest ideas, that can find very interesting applications
\cite{r}, is the so called Ghost Imaging. In summary, the idea at
the basis of this protocol is that whether one disposes of two
noise-correlated light beams, one crossing an object to be imaged
and then detected by a bucket detector without any spatial
resolution, the other addressed to a spatial resolving detector (as
a CCD camera), then the image of the object can be reconstructed by
considering correlations between the two measurements.

This protocol was firstly predicted \cite{gh1} and then demonstrated
with twin beams \cite{gh2} and later with thermal light
\cite{gh31,gh32,gh33,gh34} (with only a smaller visibility).

Recently, it was shown that the use of higher order correlation
functions can improve the visibility of ghost imaging \cite{3}.

A first experiment in this sense was realized in \cite{m}. This
interesting result, nevertheless, was valid only in the
approximation of very high light flux where $G^{(3)}$ was
reconstructed by a $G^{(2)}$ measurement and was realized with a
"fake" mask (i.e. by artificially blackening some pixels of the
camera via software). Very recently a second experiment appeared
where a real object (a double slit) was imaged through a scan that
allowed to reconstruct $G^{(3)}$ \cite{s}.

Here we present our preliminary results toward third order
correlation Ghost Imaging reconstruction of a real object observed
through a CCD camera and compare the achieved results with second
order ghost imaging ones.

\section{Theory}
\begin{figure}[!t]
\begin{center}
\includegraphics[width=12 cm, height=10 cm, bb=0 0 1200 800]{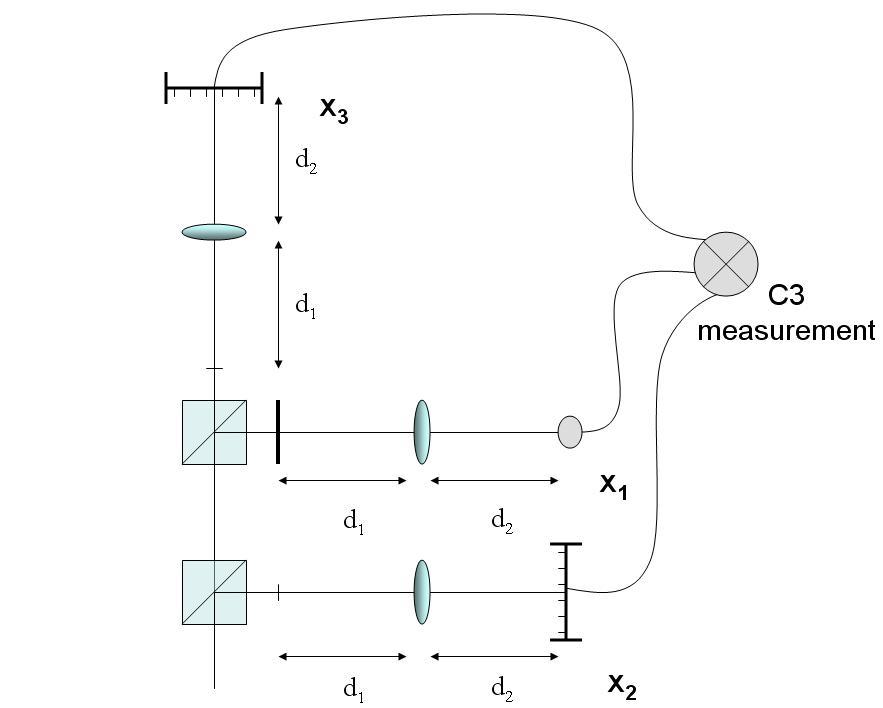}
\caption{Third order imaging scheme.  There are two reference arms
and a test arm. Distances $d_1$ and $d_2$ have to satisfy the
imaging condition. The detectors on the reference arms are spatially
resolving while in the test arm a bucket detector is present.
}\label{scheme}
\end{center}
\end{figure}

The imaging scheme (Fig. \ref{scheme}) counts three arms, two
reference arms and a test arm. The unknown object is inserted in the
latter. Each distinct imaging system is characterized by its impulse
response function $h_i(x_i^{'},x_i)$, where $i=1,2,3$, $x_i^{'}$ are
the starting transverse coordinates and $x_i$ are the coordinates at
the detection plane. Hence, considering a field $E_i(x_i^{'})$, its
value at the detection plane is given by:
\begin{equation}\label{field}
E_i(x_i)=\int dx_i^{'}h_i(x_i^{'},x_i)E_i(x_i^{'})
\end{equation}

In our scheme the subscript $i=1$ refers to the test arm.

The detectors in the reference arms are spatially resolving
detectors and record the intensity distribution $I_2(x_2)$ and
$I_3(x_3)$ in the pixel $x_2$ and $x_3$ respectively. Information
about the object is obtained by measuring the (normalized) second
order correlation coefficient $c2$ in the case of second order ghost
imaging and the (normalized) third order coefficient $c3$ in the
case of third order ghost imaging, defined as:
\begin{equation}\label{C2}
c2(x_1,x_j)=\frac{<I_1(x_1)I_j(x_j)>-<I_1(x_1)><I_j(x_j)>}{\sqrt{\mu_2(x_1)}\sqrt{\mu_2(x_j)}}
\end{equation}
with $j=2,3$ and:
\begin{equation}\label{C3}
c3(x_1,x_2,
x_3)=\frac{<(I_1(x_1)-<I_1(x_1)>)(I_2(x_2)-<I_2(x_2)>)(I_3(x_3)-<I_3(x_3)>)>}{\sqrt[3]{\mu_3(x_1)}\sqrt[3]{\mu_3(x_2)}\sqrt[3]{\mu_3(x_3)}}
\end{equation}
where $<I_i(x_i)>=<E_i^{*}(x_i)E_i(x_i)>$ is the mean intensity of
the i-th beam. In formula (\ref{C2}) we define:
\begin{equation}\label{d2}
\mu_2(x_i)=<(I_i(x_i)-<I_i(x_i)>)^2>
\end{equation}
as the second central moment, while in (\ref{C3})
\begin{equation}\label{d3}
\mu_3(x_i)=<(I_i(x_i)-<I_i(x_i)>)^3>
\end{equation}
is the third central moment. Symbols $<..>$ refers to temporal
averages. As a matter of fact it is the correlation between the
beams which is responsible of the image formation in the ghost
imaging process \cite{l,ivo}, for this reason we considered the
(normalized) correlation coefficient c2 and c3 instead of the
$G^{(2)}$ and $G^{(3)}$ functions. In fact, this is helpful as it
removes unwanted effects related to the unbalancing of the intensity
or of the fluctuations between the beams which may somehow hide the
correlations of interest.

\section{Experimental set-up}
\begin{figure}[!t]
\begin{center}
\includegraphics[width=12 cm, height=10 cm, bb=0 0 1200 800]{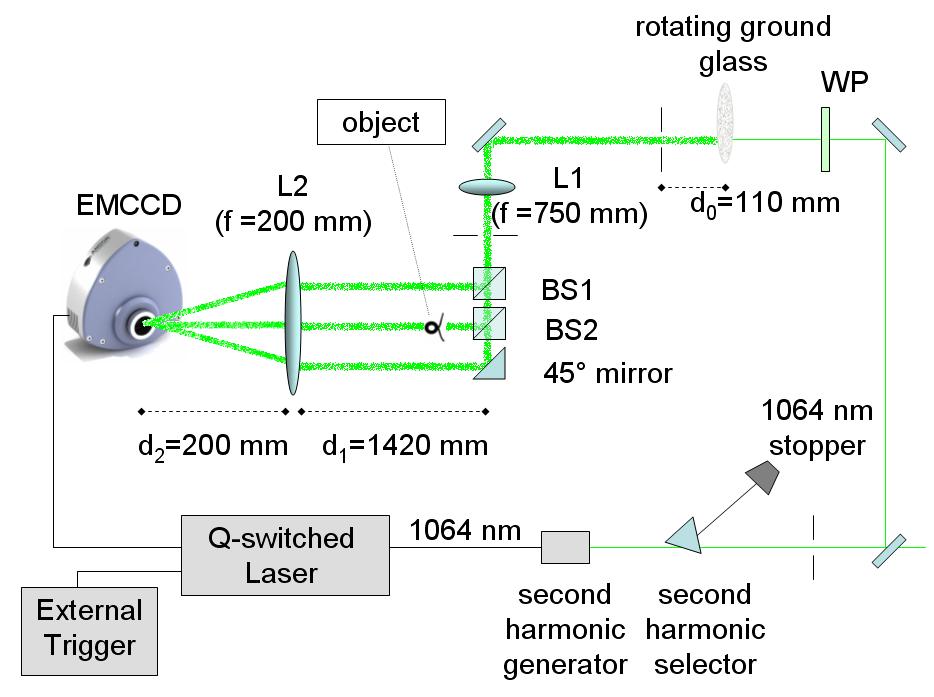}
\caption{Experimental set-up.}\label{set_up}
\end{center}
\end{figure}

In our experiment (Fig. (\ref{set_up})), the imaging system consists
of three classical correlated thermal light beams. The thermal field
is obtained by passing a coherent source through a random media. The
source is a passively Q-switched microchip laser with a pulse
duration of about 1 ns and a maximum average power of 1 W; this
source together with a frequency doubler (the second harmonic
generator in the figure) produces a 532 nm beam. The prism deflects
the first harmonic (1064 nm) beam to a beam stopper. The 532 nm
source is scattered by a rotating ground glass disk (the random
media) in order to obtain the pseudo-thermal field. The beam is then
split in three beams  by means of two beam splitters (a polarizing
beam splitter BS1 and a non-polarizing beam splitter BS2) and a
$45^\circ$ mirror and sent to a EMCCD camera.  The presence of the
half waveplate WP on the path of the beam before the rotating disk,
together with the polarizing beam splitter, allows to balance the
intensities on the three arms.

 The dimension of the coherence areas of the multimode thermal
beams are set by the lens L1 of focal length $f=750$ mm. The lens is
put in a $f-f$ configuration respect to the pinhole and to the
object in order to observe the far field speckle pattern at the
plane with the object. In this way, a single transverse wavevector
$q$ is associated to a single point $ x = (\lambda f / 2 \pi)q$. The
pinhole is at distance $d_0=110$ mm from the rotating ground glass.
 As a matter of fact, the dimension of the coherence areas is also proportional, for a given
wavelength, to the distance between the ground glass and the pinhole
and inversely to the dimension of the pinhole. A second pinhole
after L1 determines the dimension of the source.

The lens L2 before the camera is used to image the beams on the CCD.
A magnification occurs in order to image the whole area of the three
beams on the $658$ x $498$ pixels sensitive area of the CCD, that is
$6,58$ mm x $4,98$ mm since the pixel area is $10$ $\mu$m x $10$
$\mu$m. Distances $d_1=1420$ mm and $d_2 = 200$ mm are chosen in
order to obtain a proper demagnification.

 The camera is triggered with the
pulses of the laser. The repetition rate of the laser is externally
set to $500$ Hz, while the exposure time of the camera is of $2$ ms.
Since the pulse duration is no longer than $1$ ns, each frame taken
with the CCD contains only one pulse of the laser. The ratio between
the repetition rate of the camera and the frequency of rotation of
the disk avoids a periodic speckle pattern detection.  Moreover, the
disk has to rotate slowly enough to be considered stationary with
respect to the duration of the pulse of the laser.

The object is an iron curled wire with a diameter of 1.5 mm.

\section{Toward third-order ghost imaging: experimental results}

\begin{figure}[!t]
\begin{center}
\includegraphics[width=12 cm, height=10 cm, bb=0 0 1200 800]{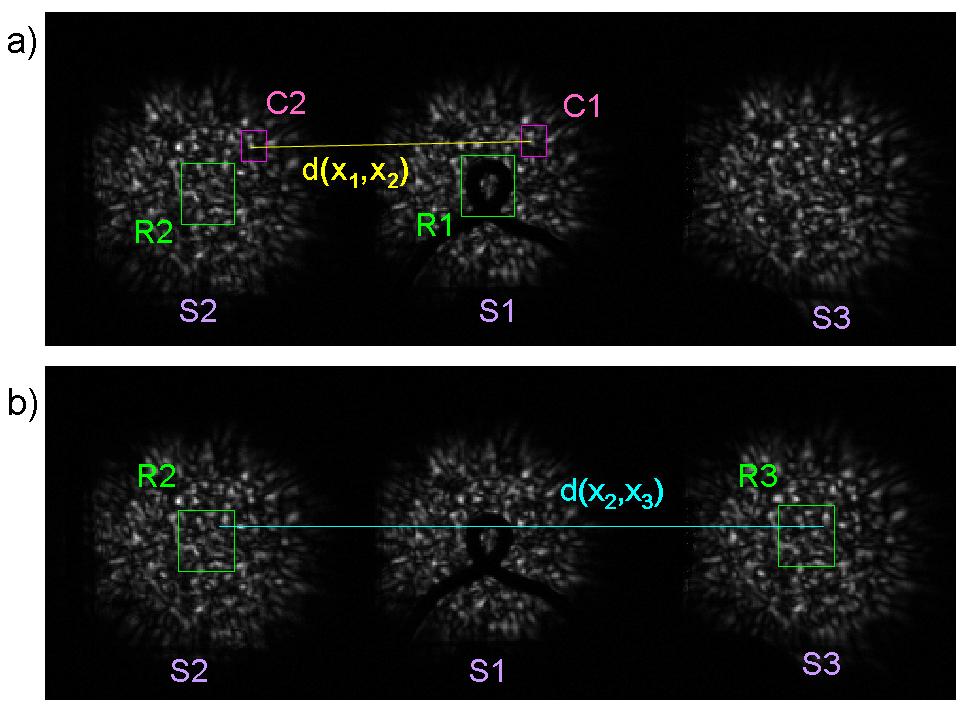}
\caption{The image of the three thermal beams from a single shot of
the laser a)Correlated regions in arm 1 and 2. b)Correlated regions
in arm 2 and 3.}\label{frame}
\end{center}
\end{figure}

Fig. (\ref{frame}) reports the image obtained from a single shot of
the laser. We can appreciate the speckle structure and the
correlations in its intensity on the three thermal beams. The object
is in the central arm, named $S_1$; $S_2$ and $S_3$ denotes the
reference arms. In order to retrieve information about the
correlations, the first step is to determine with high precision the
regions $R_2$ and $R_3$ correlated to the one with the object, $R_1$
(Fig. (\ref{frame}) a)). Due to the presence of the object, we first
determine the distance $d_{1,2}$ between $R_1$ and $R_2$ using a
region $C_1$ without the object and finding the correlated one,
$C_2$ in the arm 2. Starting from a region $C_{2,0}$ of $S_2$ we
displace the region in horizontal and vertical directions, pixel by
pixel, in order to find the maximum of the normalized second order
correlation coefficient (\ref{C2}) for a frame $k$:

\begin{equation}\label{C2p}
c2^{k} =
\frac{<I^{(k)}_1(x_1)I^{(k)}_2(x_{2,0}+D)>-<I^{(k)}_1(x_1)><I^{(k)}_2(x_{2,0}+D)>}{\sqrt{\mu_2(x_1)}\sqrt{\mu_2(x_{2,0}+D)}}
\end{equation}

where $I^{(k)}_1(x_1)$ and $I^{(k)}_2(x_{2,0})$ are the intensities
of the frame $k$ for the region $C_1$ and $C_{2,0}$ respectively.
$D$ is the displacement vector  in pixel. In Eq. (\ref{C2p}) the
symbol $<..>$ refers to spatial averages over the pixels of the
selected regions. The distance between the correlated regions is
computed as: $d_{1,2}=|x_{1}-x_2|$, with $x_2=x_{2,0}+D_{max}$,
where $D_{max}$ correspond to the displacement to be applied in
order to reach the maximum of the correlation. Finally we determine
the region $R_3$ correlated to $R_2$ using the same procedure (Fig.
\ref{frame} b)).

\begin{figure}[!t]
\begin{center}
\includegraphics[width=12 cm, height=8 cm, bb=0 0 1200 800]{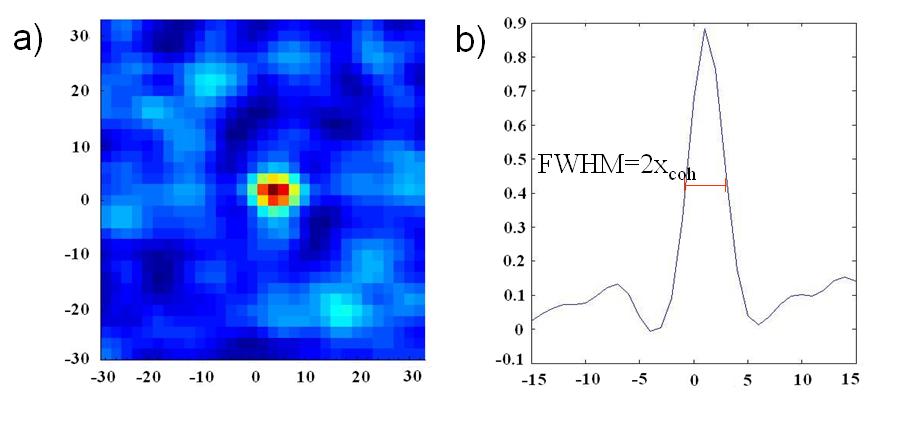}
\caption{a) The correlation coefficient as a function of the
displacement vector $D$ b) The peak of correlation. Its FWHM
represents the diameter of the speckle}\label{speckle}
\end{center}
\end{figure}

 The correlation coefficient as a function of the displacement
vector $D$ is depicted in Fig. (\ref{speckle}) a); the part b) of
the figure reports the section of the peak of correlation in the
horizontal direction. The reported correlation between the regions
is quite high ($\sim 0.88$), denoting a low level of losses in the
optical path and low noise. The FWHM of the peak is a good
estimation of the size of a speckle; in our experiment we generate
speckles of radius $x_{coh}\thickapprox 3$ pixels $= 30$ $\mu$m.

To perform ghost imaging, we use a bucket detector on arm 1. For
each frame $k$ taken with the camera, the measured quantity in arm 1
is then the integral over $x_1$: $I_1^{(k)}=\int dx_1
I^{(k)}_{1}(x_1)$. In practice, it corresponds to the sum over the
pixels of the region $R_1$ with the object. The intensity
distributions of the correlated regions on the other arms are
registered as an array of intensity values $I^{(k)}_j(x_j)$.

\begin{figure}[!t]
\begin{center}
\includegraphics[width=12 cm, height=7 cm, bb=0 0 1200 800]{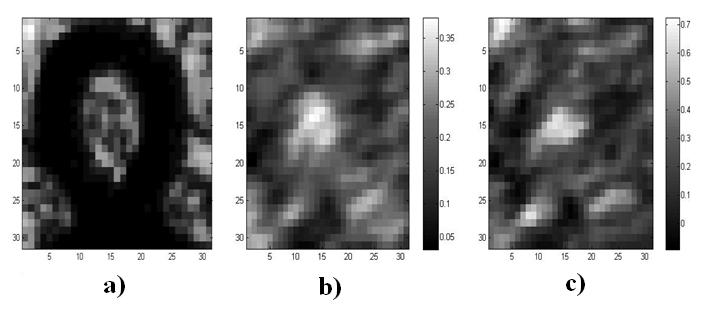}
\caption{a) The image of the real object b) The second order ghost
image c) The third order ghost image}\label{g2}
\end{center}
\end{figure}

We reconstruct normalized second and third order correlation
functions c2 and c3 as the ghost images of our real object. For c2
we used  arm 2 and arm 3 alternatively as reference. Fig. (\ref{g2})
shows our preliminary results. Part a) of the figure reports the
single frame image of the object in the region $R_1$, while part b)
shows one of the two reconstructed images by means of c2. Finally,
the c) part is the third order ghost imaging of the object. The
correlation coefficients are computed averaging over $N_f=400$
frames. The reconstructed images are evident even if the resolution
is quite low. This can be understood once you consider that the
speckles have a diameter of 6 speckles, i.e. a dimension comparable
to the one of the object (i.e. tens of speckles according to Fig.
(\ref{g2}) a))


As a figure of merit to estimate the improvement of third order
ghost imaging respect to second order one we chose the visibility of
the reconstructed images, here defined as:
\begin{equation}\label{C2p}
V_j =\frac{cj_{back}-cj_{obj}}{cj_{back}+cj{x_{obj}}}
\end{equation}
with $j=2,3$, where $cj_{back}$ is the value of the correlation
coefficient averaged over a region without the object and $cj_{obj}$
is computed averaging over a region with the reconstructed object.

We obtain  $V_2=0.2301 \pm 0.0008$ (the result is independent by the
choice of arm 2 or 3 as reference) and $V_3=0.269 \pm 0.006$. It is
clear that there is an improvement in the value of the visibility of
the reconstructed image by means of third order correlations. It was
theoretically predicted by Bai et al. \cite{3} that the existence of
additional correlation parts, namely correlations between the two
reference detectors and between the three detectors, leads to an
improvement of the visibility. Our experimentally measured increment
in the value of $V_3$ with respect to $V_2$ can be compared to their
theoretical predictions. Moreover, the gap remains stable even if we
use different set of images. We also observe that the number of
frames used to reconstruct the image is sufficient to reach the
maximum gap. An increment of the number of frames do not lead to an
improvement of the ghost imaging performances. In any case, we want
to stress that the reported data are a preliminary experimental
demonstration of the improvement in the visibility and that a more
complete analysis is postponed to a subsequent paper.

\section{Conclusions}
In this paper we have presented a realization of third order ghost
imaging of a real object exploiting a CCD camera. We also show an
improvement in the visibility of third order ghost imaging respect
to second order one. Having overcome some limits of previous
experiments, it represents a relevant step toward real applications
of ghost imaging \cite{r}.

\section*{Acknowledgments}

We thank M. Chekhova for useful discussions.

\end{document}